\documentclass[%
reprint,
amsmath,amssymb, 
aps,
prb
]{revtex4-2}
\usepackage{graphicx, color}
\usepackage{dcolumn}
\usepackage{bm}
\usepackage[colorlinks,urlcolor=blue,linkcolor=blue,anchorcolor=blue,citecolor=blue]{hyperref}

\begin{document}
\preprint{APS/123-QED}
\title{Twist-induced magnetic topological phase transition in stacked altermagnetic CrO}
\author{Zi-Hao Ding$^{1}$}
\author{Ze-Feng Gao$^{1}$}
\author{Xiang-Hua Kong$^{2}$}
\author{Peng-Jie Guo$^{1}$}
\email{guopengjie@ruc.edu.cn}
\author{Zhong-Yi Lu$^{1,3}$}
\email{zlu@ruc.edu.cn}
\affiliation{1. School of Physics and Key Laboratory of Quantum State Construction and Manipulation (Ministry of Education), Renmin University of China, Beijing 100872, China}
\affiliation{2. College of Physics and Optoelectronic Engineering, Shenzhen University, Shenzhen 518060,China}
\affiliation{3. Hefei National Laboratory, Hefei 230088, China}
\date{\today}

\begin{abstract}
Interlayer twisting offers a geometric route to controlling electronic states, but whether it can simultaneously reconstruct magnetic symmetry and band topology remains unclear. Here, based on symmetry analysis and first-principles calculations, we show that commensurate twisting drives magnetic topological phase transitions in stacked bilayer CrO. In particular, it transforms an antiferromagnetic Dirac semimetal into either a $d$-wave altermagnetic bipolarized Weyl semimetal or an unconventional compensated magnetic Weyl semimetal. A key result is that the Weyl points in the $d$-wave altermagnetic phase lie at generic $k$ points in the Brillouin zone and are protected by the spin symmetry $\left\{ C_2 T||C_{2z} T\right\}$. This sharply contrasts with conventional two-dimensional Weyl semimetals, where Weyl points are typically protected by mirror or rotational symmetries and thus pinned to high-symmetry lines. We further show that commensurate twisting preserves the spin symmetry $\left\{ C_2 T||C_{2z} T\right\}$, making the Weyl phase a robust consequence of twisting rather than a fine-tuned feature of a specific angle. Our work establishes a symmetry-based route to engineering magnetic topological phases in twisted two-dimensional materials.
\end{abstract}

\maketitle

\textit{Introduction.}
Topological semimetals have attracted sustained interest because of their nontrivial band topology and symmetry-protected band crossings near the Fermi level \cite{burkov2016,weng2016,RevModPhys.90.015001, zou2019,RevModPhys.93.025002,bernevig2022}. They host a variety of remarkable phenomena, including ultrahigh mobility \cite{ultrahigh}, negative magnetoresistance \cite{Chiralanomalyandnegative,NegativeMagnetoresistance}, chiral anomaly \cite{chiralanomaly,Chiralanomalyandnegative}, and unconventional Hall responses \cite{nsr}. Yet most known topological semimetals realize only a single type of band topology \cite{PhysRevX.5.031013,PhysRevX.5.011029,Science349,PhysRevB.85.195320,PhysRevLett.108.140405,Science353,PhysRevX.6.031003,adma.201906046,PhysRevLett.115.036806,PhysRevB.92.045108,fang2016topological,PhysRevLett.117.096401}. It is therefore highly desirable to identify controllable routes for driving topological phase transitions and coupling topology to other quantum orders within a single material platform.

Altermagnetism has recently emerged as a particularly promising setting for this purpose \cite{JPSJ.88.123702,sciadv.aaz8809,PhysRevB.102.014422,mazin2021,PhysRevX.12.040002,PhysRevX.12.031042,PhysRevX.12.040501,adfm.202409327,song2025altermagnets,2026symmetry}. As a distinct magnetic phase, it features both zero net magnetization and momentum-dependent spin splitting in reciprocal space. Altermagnets thus unite key features of antiferromagnets and ferromagnets, making them natural candidates for realizing topological phases with unconventional magnetic structure \cite{guo2023quantum, MirrorChern,nodaltonodeless,tan2024,CVHE,squ,feng2025,TAN2026,Quantizedspin,GdAlSi,CrSb}.

A major open question is how to controllably engineer altermagnetic order while simultaneously tuning the associated band topology. Moir\'e systems offer an appealing answer \cite{carr2020electronic,andrei2021marvels,du2021engineering,kennes2021moire,mak2022semiconductor}. By introducing a relative twist angle between stacked two-dimensional layers, they provide a highly tunable geometric degree of freedom capable of reshaping electronic structure without chemical modification \cite{PhysRevB.95.075420}. Recent studies have shown that twisting can induce altermagnetism in conventional magnetic van der Waals bilayers \cite{PhysRevLett.133.206702,PhysRevLett.130.046401,PhysRevMaterials.8.L051401,PhysRevB.110.174410}, while twist engineering can also generate diverse topological phases in few-layer materials \cite{li2021quantum,zeng2023thermodynamic,tian2025real}. However, twist-induced altermagnetism and twist-controlled topology have largely been treated separately.

This separation is especially significant for two-dimensional Weyl semimetals. In conventional cases, Weyl points are typically protected by mirror or rotational symmetries and are therefore confined to high-symmetry lines in the Brillouin zone. Whether twisting can instead generate a two-dimensional Weyl phase with symmetry-protected Weyl points at generic momenta remains unknown.

In this Letter, we answer this question in stacked bilayer CrO by combining symmetry analysis and first-principles calculations. We show that the high-symmetry AA, AB, and AC stackings all transform monolayer $d$-wave altermagnetic CrO from a bipolarized Weyl semimetal \cite{tan2024} into an antiferromagnetic Dirac semimetal, with AB stacking being the most stable. We then show that two commensurate twist angles, $36.87^\circ$ and $22.62^\circ$, drive the AB bilayer back into a $d$-wave altermagnetic bipolarized Weyl semimetal. In this phase, four pairs of Weyl points appear at generic $k$ points and are protected by the spin symmetry $\left\{ C_2 T||C_{2z} T\right\}$. More importantly, this spin symmetry is preserved under commensurate twisting, implying that the transition from an antiferromagnetic Dirac semimetal to a $d$-wave altermagnetic bipolarized Weyl semimetal is a robust consequence of commensurate twisting rather than a fine-tuned feature of a particular angle. We further find that AA stacking behaves similarly, while AC stacking evolves into an unconventional compensated magnetic Weyl semimetal \cite{GUO_2025,Hou_2025}. These results establish interlayer twisting as a symmetry-guided route to magnetic topological phase transitions in two-dimensional materials.

\begin{figure}[t]
	\centering
	\includegraphics[width=8.5 cm]{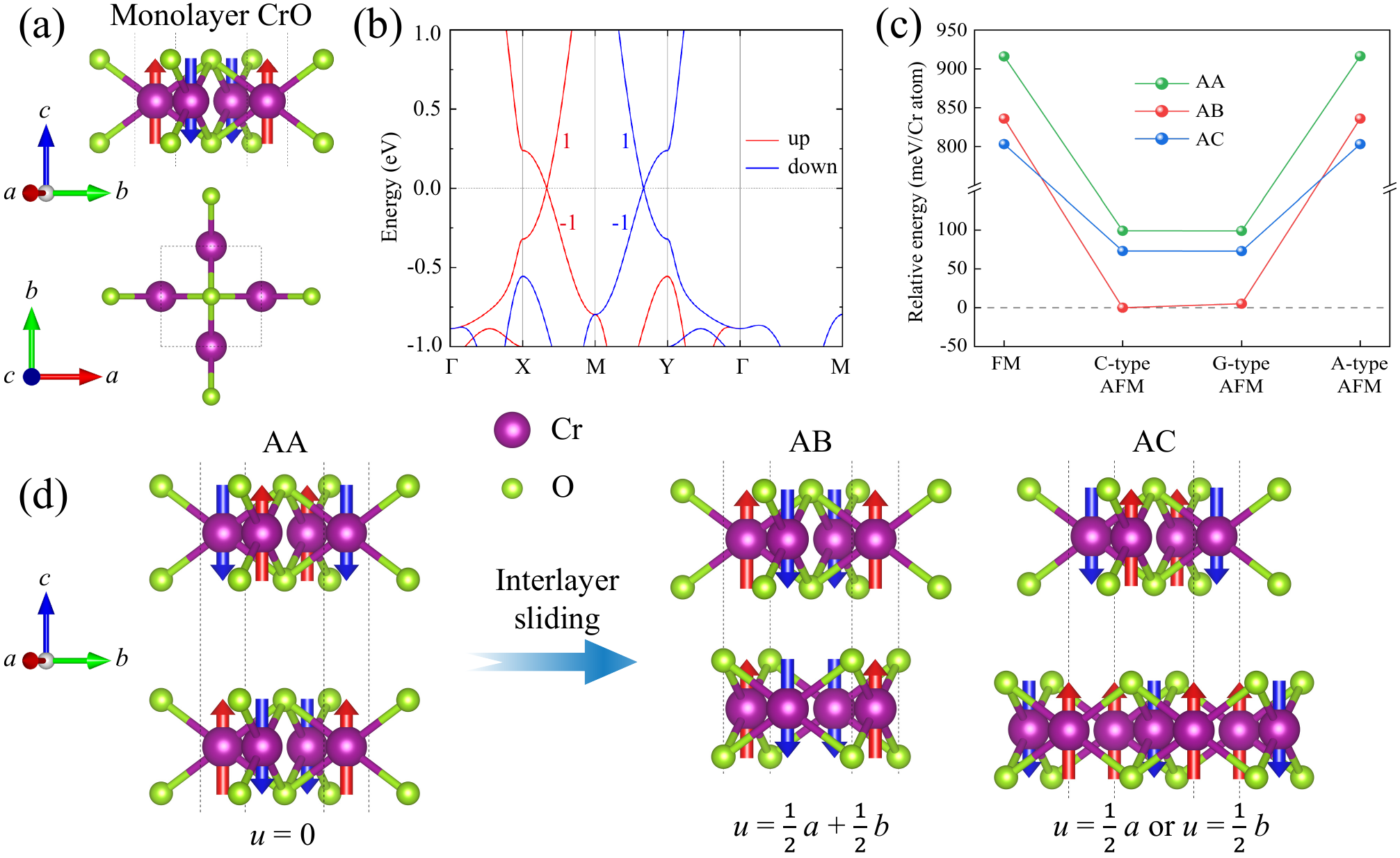}
	\caption{(a) Crystal structure and ground-state magnetic configuration of monolayer CrO. Cr and O atoms are colored purple and green, respectively, while red and blue arrows indicate opposite orientations of the local magnetic moments. (b) Electronic band structure of monolayer CrO calculated without spin-orbit coupling (SOC). The eigenvalues of the mirror symmetry $\left\{ E|| M_{x} \right\}$ for the two crossing bands along the X-M direction are denoted by +1 (red) and -1 (red), while those of $\left\{ E|| M_{y} \right\}$ along the M-Y direction are +1 (blue) and -1 (blue). (c) Calculated relative energies of different magnetic configurations for three stackings. Here, the C-type AFM of AB stacking is chosen as the energy reference. (d) Crystal structures and ground-state magnetic configurations of AA, AB, and AC stackings of bilayer CrO and their sliding vector $\boldsymbol{u}$.}
	\label{fig1}
\end{figure}


\textit{Results and analysis.}
The monolayer CrO has a tetragonal lattice with space group $P4/mmm$ (No.~123; point group $D_{4h}$), generated by $C_{4z}$, $C_{2x}$  and $I$. The crystal and magnetic structures are shown in Fig.~\ref{fig1}(a). Each primitive unit cell contains four atoms (two Cr and two O). The two opposite spin sublattices are related by $\left\{ C_2|| C_{4z} \right\}$ symmetry instead of $\left\{ C_2||I \right\}$ symmetry. Meanwhile, the two nonmagnetic O atoms break $\left\{ C_2||\tau \right\}$ symmetry, rendering monolayer CrO a $d$-wave altermagnet. The optimized lattice constants are $a = b = 3.324\ \text{\AA}$, which are consistent with the previous study \cite{guo2023quantum,APL-CrO}. We further calculate the band structure along high-symmetry lines without spin-orbit coupling (SOC). As shown in Fig.~\ref{fig1}(b), the band structure is spin-degenerate along the $\Gamma$-M direction due to $\left\{ C_2|| M_{xy} \right\}$ symmetry, while it splits in other directions. The emergence of Weyl points is clearly visible along the X-M and M-Y directions, protected by $\left\{ E|| M_{x} \right\}$ and $\left\{ E|| M_{y} \right\}$ symmetries, respectively. Thus, monolayer CrO is identified as a $d$-wave altermagnetic bipolarized Weyl semimetal. The consistency with previous studies \cite{guo2023quantum,APL-CrO} confirms the validity of our calculations.

If we assemble two monolayer structures vertically into a bilayer structure, different stacking patterns will endow the system with different symmetries, offering the possibility of emergent properties far beyond those of monolayer. Here, we consider three high-symmetry stackings, AA, AB, and AC. As shown in Fig.~\ref{fig1}(d), the AA stacking is constructed by vertically aligning two monolayers along the $z$ direction. The AB and AC stackings can be obtained by interlayer sliding of the AA stacking. In the AB stacking, the bottom-layer CrO slides by $\frac{1}{2}\boldsymbol{a} + \frac{1}{2}\boldsymbol{b}$ relative to the top-layer CrO, while in the AC stacking, the bottom-layer CrO slides by $\frac{1}{2}\boldsymbol{a}$ or $\frac{1}{2}\boldsymbol{b}$ relative to the top-layer CrO. In our calculations, several possible magnetic configurations are considered, including FM, C-type AFM, G-type AFM, and A-type AFM, which are shown in Fig.~S1 of the Supplemental Material (SM) \cite{SM}. Fig.~\ref{fig1}(c) shows the relative energies of different magnetic configurations for AA, AB, and AC stackings. For AA and AC stackings, the energy of G-type AFM is 0.095 and 0.22 meV/Cr lower than that of C-type AFM, respectively, indicating that G-type AFM is the magnetic ground state. For the AB stacking, C-type AFM is the magnetic ground state, with an energy 5.201 meV/Cr lower than that of G-type AFM. The ground-state magnetic configurations of the three stackings are shown in Fig.~\ref{fig1}(d). Moreover, the AB stacking is the most stable structure, and the AC stacking is the substable structure with an energy 72.853 meV/Cr higher than that of the AB stacking. The AA stacking is the most unstable structure with an energy 98.958 meV/Cr higher than that of the AB stacking. Therefore, in the main text, we focus on the most stable structure, the AB stacking, while the results for these other two stackings are provided in the SM \cite{SM}.

\begin{figure}[htbp]
	\centering
	\includegraphics[width=8.5cm]{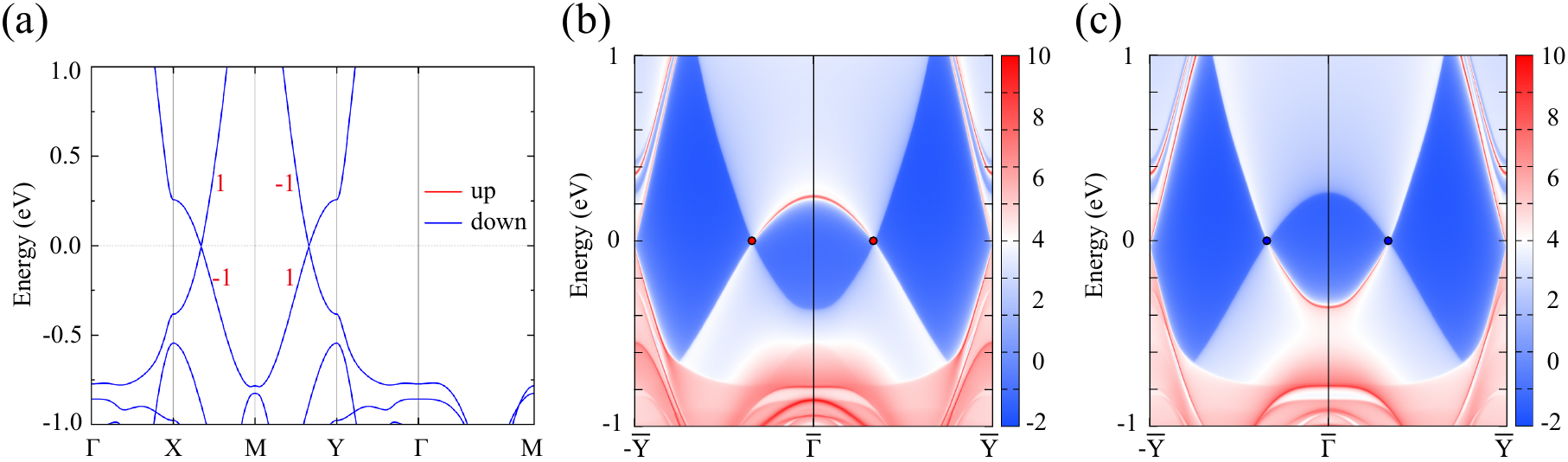}
	\caption{(a) Electronic band structure of AB stacking calculated without SOC. The eigenvalues of the mirror symmetry $\left\{ E|| M_{x} \right\}$ for the two crossing spin-up bands along the X-M direction are denoted by +1 (red) and -1 (red), while those of $\left\{ E|| M_{y} \right\}$ along the M-Y direction are +1 (red) and -1 (red). (b), (c) Spin-up and spin-down spectral functions for AB stacking projected onto the (100) termination, respectively. The color scales represent the local density of states, with warmer colors indicating larger values. Red and blue dots mark the spin-up and spin-down Weyl points, respectively.}
	\label{fig2}
\end{figure}

The AB stacking has a tetragonal lattice with space group $P4/nmm$ (No.~129; point group $D_{4h}$), generated by $C_{4z}(\frac{1}{2},\frac{1}{2},0)$, $C_{2x}(\frac{1}{2},\frac{1}{2},0)$, and $I(\frac{1}{2},\frac{1}{2},0)$. The optimized lattice constants are $a = b = 3.299\ \text{\AA}$. Cr ions with opposite magnetic moments are connected via the $I(\frac{1}{2},\frac{1}{2},0)$ symmetry, rendering the AB stacking a conventional collinear antiferromagnet. As shown in Fig.~\ref{fig2}(a), without spin-orbit coupling (SOC), the calculated band structure of AB stacking reveals a gapless semimetallic nature, featuring four linear band crossings near the Fermi level. The orbital-projected band structure in Fig.~S3(b) further indicates that the bands near the Fermi level mainly originate from the $d_{x^2 - y^2}$ orbital of Cr. For the spin-up bands, the band crossing along the X-M direction involves two states near the Fermi level belonging to different irreducible representations, with $\left\{ E|| M_{x} \right\}$ mirror eigenvalues of +1 and -1; meanwhile, the band crossing along the M-Y direction corresponds to $\left\{ E|| M_{y} \right\}$ mirror eigenvalues of -1 and +1. Moreover, due to the $\left\{ T||IT(\frac{1}{2},\frac{1}{2},0) \right\}$ symmetry, all bands remain spin-degenerate. Therefore, the AB stacking is an antiferromagnetic Dirac semimetal protected by mirror symmetry.

Furthermore, we investigate the magnetocrystalline anisotropy of this bilayer CrO system by comparing the total energies under different magnetization directions. The results show that the energies for in-plane magnetization directions are identical to each other and are 0.003 meV/Cr lower than that for the [001] direction. This weak magnetic anisotropy implies that the magnetism in AB stacking is very soft, allowing an external magnetic field to easily switch its magnetization direction. When the magnetization is along [100] (the easy axis), an energy gap of approximately 1 meV opens at the Dirac point along the X-M direction, while a small gap of about 3 meV opens at the Dirac point along the M-Y direction (as shown in Fig.~S3(a)). Clearly, the SOC is so weak in this system that it can be neglected.

Nontrivial topological edge states with Fermi arcs are commonly considered to be a fingerprint of Dirac semimetals. Therefore, we use the surface Green's function technique to calculate the edge states of AB stacking projected onto the (100) termination, as shown in Figs.~\ref{fig2}(b) and \ref{fig2}(c). It is clear that Fermi arc states emerge on the edge, and they connect the two Dirac points. 

After characterizing the untwisted bilayer, we next investigate twisted bilayer CrO. The commensurate supercells of twisted bilayer CrO are constructed using the coincidence-site lattice method \cite{carr2020electronic,PhysRevLett.99.256802,PhysRevB.85.195458,koda2016coincidence} (The detailed construction procedure is provided in the SM \cite{SM}). Due to limited computational resources, we focus on a large twist angle $\theta = 36.87^\circ$ with $(M,N)=(2,1)$, whose 40-atom supercell allows a detailed characterization.

\begin{figure}[htbp]
	\centering
	\includegraphics[width=8.5cm]{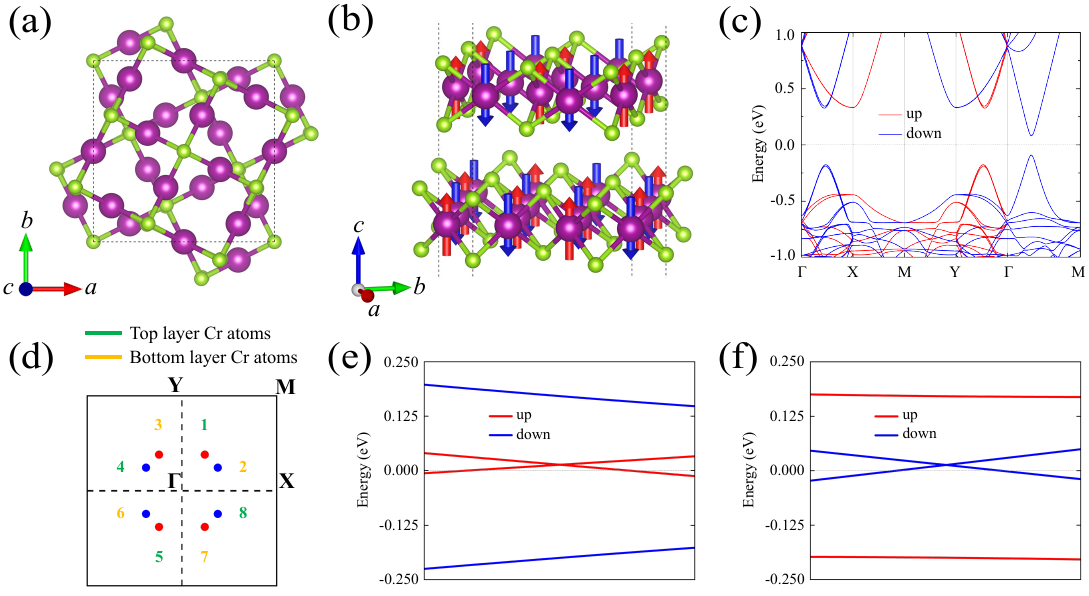}
	\caption{(a) Crystal structure of the 40-atom moiré supercell and (b) ground-state magnetic configuration for the twisted bilayer of AB stacking (tb-AB). (c) Electronic band structure of tb-AB calculated without SOC. (d) Distribution of Weyl points in the Brillouin zone. Red and blue dots mark the spin-up and spin-down Weyl points, respectively, while dark green and orange indicate the contributions from Cr atoms in the top and bottom layers, respectively. (e), (f) Enlarged views of the band structure without SOC around Weyl points 1 and 2 near the Fermi level, respectively.}
	\label{fig3}
\end{figure}

The twisted bilayer of AB stacking (tb-AB) has a tetragonal lattice with space group $P42_{1}2$ (No.~90; point group $D_{4}$), generated by $C_{4z}(\frac{1}{2},\frac{1}{2},0)$, $C_{2x}(\frac{1}{2},\frac{1}{2},0)$. Its crystal structure is shown in Fig.~\ref{fig3}(a). To determine the magnetic ground state of tb-AB, we consider the C-type and G-type AFM configurations shown in Figs.~S2(e) and S2(f). Our calculations reveal that tb-AB possesses the G-type AFM configuration, as depicted in Fig.~\ref{fig3}(b). This configuration is energetically more stable than the C-type AFM by 0.068 meV/Cr. The optimized lattice constants are $a = b = 7.353\ \text{\AA}$. In each layer, the two opposite spin sublattices are connected by $\left\{ C_2|| C_{4z} \right\}$ symmetry, and the O atoms break $\left\{ C_2||\tau \right\}$ symmetry. The introduction of the twist angle breaks the original $\left\{ C_2|| I(\frac{1}{2},\frac{1}{2},0) \right\}$ symmetry in the AB stacking, thereby transforming the system from a conventional antiferromagnet into a $d$-wave altermagnet. Interestingly, our calculations indicate that the magnetic easy-axis of tb-AB is out-of-plane, with an energy 0.007 meV/Cr lower than that of the in-plane magnetization directions. That is to say, after twisting, the magnetic easy-axis of AB stacking transitions from in-plane to out-of-plane. From an application perspective, this perpendicular magnetic anisotropy is significant for spintronic devices, especially for the multilayer integration. 

According to the symmetry analysis, the corresponding band structure is also spin-degenerate along the $\Gamma$-M direction due to $\left\{ C_2|| C_{2xy} \right\}$ symmetry but splits in other directions. 
Fig.~\ref{fig3}(c) shows the band structure of tb-AB along the high-symmetry directions, which is consistent with our symmetry analysis. Unlike monolayer and AB-stacked bilayer CrO, tb-AB exhibits $d$-wave altermagnetic semiconducting behavior along the high-symmetry directions. Our previous work \cite{squ} shows that the $\left\{C_2T||IT\right\}$ spin symmetry protects Weyl points at generic positions in the Brillouin zone (BZ). However, tb-AB breaks inversion symmetry and therefore does not possess the $\left\{C_2T||IT\right\}$ spin symmetry. Interestingly, in the tb-AB system, the $\left\{ C_2 T||C_{2z} T\right\}$ symmetry exists at every $k$ point in the BZ, and this symmetry can also protect Weyl points at generic positions in the BZ of magnetic systems. Consequently, tb-AB may be an altermagnetic bipolarized Weyl semimetal protected by the $\left\{ C_2 T||C_{2z} T\right\}$ spin symmetry.

Furthermore, we calculate the energy gap between the conduction and valence bands across the entire BZ and identify eight Weyl points, four in the spin-up channel and four in the spin-down channel, as shown in Fig.~\ref{fig3}(d). Due to the $\left\{C_2||C_{4z} \right\}$ symmetry of tb-AB, only two of these Weyl points are independent. These two independent Weyl points are marked in the band structure (Figs.~\ref{fig3}(e) and \ref{fig3}(f)). From a broader perspective, AB stacking itself is an antiferromagnetic Dirac semimetal with four Dirac points. The twisting action transforms AB stacking from a conventional antiferromagnet into an altermagnet, leading to the breaking of the effective time-reversal symmetry, namely the $IT(\frac{1}{2},\frac{1}{2},0)$ symmetry. Each Dirac point consequently converts into two Weyl points, thereby rendering tb-AB an altermagnetic bipolarized Weyl semimetal with eight Weyl points. Moreover, the layer-projected band structures show that Weyl points 1, 4, 5, and 8 are contributed by orbitals from the top-layer Cr atoms, while Weyl points 2, 3, 6, and 7 are contributed by orbitals from the bottom-layer Cr atoms. In addition, regardless of the commensurate twist angle, the spin symmetry $\left\{ C_2 T||C_{2z} T\right\}$ remains stable, indicating that the Weyl semimetal phase in tb-AB exhibits extremely strong robustness.

\begin{figure}[t]
	\centering
	\includegraphics[width=8.5 cm]{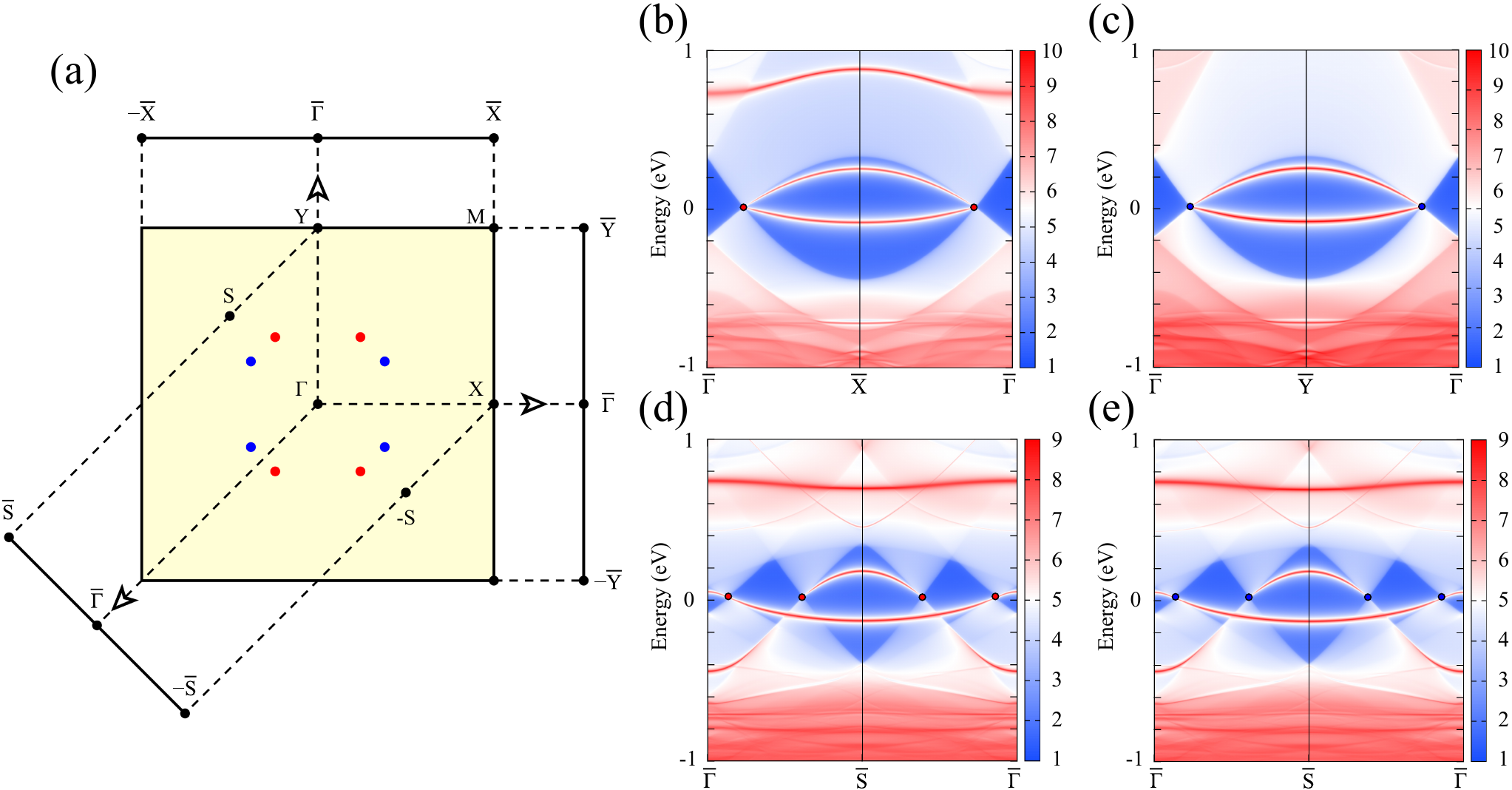}
	\caption{(a) Schematic illustration of the projections from the two-dimensional Brillouin zone of tb-AB onto the one-dimensional edge Brillouin zones associated with the (010), (100), and (110) terminations, shown at the top, right, and lower left, respectively. (b), (c) Spin-up and spin-down spectral functions for tb-AB projected onto the (010) and (100) terminations, respectively. (d), (e) Spin-up and spin-down spectral functions projected onto the (110) termination, respectively.}
	\label{fig4}
\end{figure}

Next, we employ the surface Green's function technique to calculate the spin-up channel projected onto the (010) termination, the spin-down channel projected onto the (100) termination, and each spin channel projected onto the (110) termination, with the results shown in Figs.~\ref{fig4}(a)--\ref{fig4}(e).
For the (010) and (100) terminations, the four Weyl points project onto two distinct $k$-points. Therefore, two topological edge states are observed on these terminations, connecting one pair of projected points (Figs.~\ref{fig4}(b) and \ref{fig4}(c)). In contrast, for the (110) termination, the four Weyl points project onto four different $k$-points, resulting in two topological edge states that each connect an independent pair of Weyl points (Figs.~\ref{fig4}(d) and \ref{fig4}(e)). Notably, these topological edge states span a large portion of the Brillouin zone. Such a broad edge-state distribution not only facilitates direct experimental observation using techniques such as angle-resolved photoemission spectroscopy (ARPES) but also provides an ideal platform for realizing low-dissipation topological electronic devices.

\begin{figure}[htbp]
	\centering
	\includegraphics[width=8.5cm]{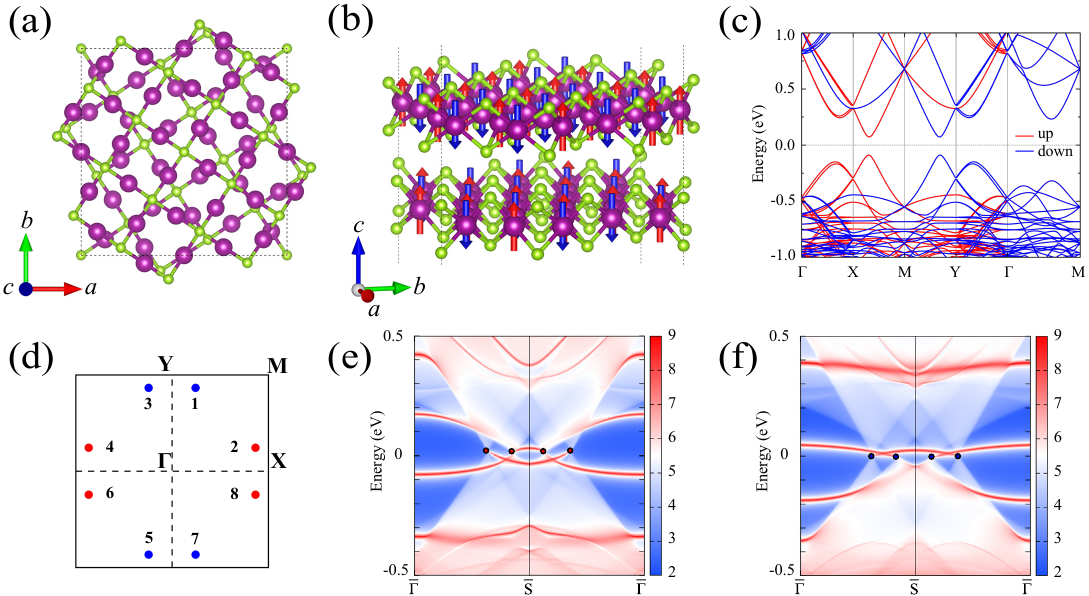}
	\caption{(a) Crystal structure of the 104-atom moiré supercell and (b) ground-state magnetic configuration for the twisted bilayer of AB stacking (tb-AB-$22.62^\circ$). (c) Electronic band structure of tb-AB-$22.62^\circ$ calculated without SOC. (d) Distribution of Weyl points in the Brillouin zone. Red and blue dots mark the spin-up and spin-down Weyl points, respectively. (e), (f) Spin-up and spin-down spectral functions projected onto the (110) termination, respectively.}
	\label{fig5}
\end{figure}

To examine whether the Weyl phase is specific to the relatively large twist angle of $36.87^\circ$, we further consider the commensurate twist angle $\theta=22.62^\circ$ with $(M,N)=(3,2)$. The corresponding
moiré supercell contains 104 atoms, as shown in Fig.~\ref{fig5}(a). In contrast to tb-AB-$36.87^\circ$, whose magnetic ground state is the G-type AFM configuration, tb-AB-$22.62^\circ$ favors the C-type AFM
configuration [Fig.~\ref{fig5}(b)], which is lower in energy than the G-type AFM configuration by 0.195 meV/Cr. Despite this change in the magnetic ground state, the spin-resolved band structure retains
pronounced momentum-dependent spin splitting [Fig.~\ref{fig5}(c)]. A Brillouin-zone-wide search identifies eight Weyl points at generic momenta, including four in each spin channel, as shown in Fig.~\ref{fig5}(d). Furthermore, the spin-resolved spectral functions projected onto the (110) termination exhibit topological edge states connecting the projected Weyl points [Figs.~\ref{fig5}(e) and \ref{fig5}(f)]. These results demonstrate that, although the twist angle modifies the relative stability of the magnetic configurations and shifts the Weyl points in momentum space, the Weyl semimetal phase persists at different commensurate twist angles. The physical origin behind this behavior is that a commensurate twist angle always drives tb-AB from a conventional antiferromagnetic state into an altermagnetic state, while preserving the $\{C_{2}T \parallel C_{2z}T\}$ spin symmetry.

Finally, for AA and AC stacking, their band structures are similar to those of AB stacking, all being antiferromagnetic Dirac semimetals (see Fig.~S4 and S5). However, after applying a twist angle of $36.87^\circ$, AA stacking transforms into a $d$-wave altermagnetic bipolarized Weyl semimetal (see Fig.~S7 and S8). The distribution of Weyl points in tb-AA is similar to that in tb-AB, except that the positions of the spin-up and spin-down Weyl points are interchanged at each corner of the BZ. On the other hand, AC stacking becomes an unconventional compensated magnetic Weyl semimetal, the two opposite spin sublattices are not related by any symmetry, leading to a complete spin-splitting of the energy bands. But its Weyl points distribution is similar to that of tb-AA.

\textit{Discussion and conclusion.}
In summary, we systematically investigate the MTPT in stacked bilayer CrO based on symmetry analysis and first-principles calculations. Monolayer CrO exhibits the characteristics of a $d$-wave altermagnetic bipolarized Weyl semimetal. By stacking two monolayers into bilayer structures, we explore three high-symmetry stackings (AA, AB, and AC) and identify their magnetic ground states. Among them, AB stacking is the most stable. All three stackings are antiferromagnetic Dirac semimetals with similar band structures. Two pairs of Dirac points are distributed along high-symmetry lines, suggesting that the stacking pattern does not significantly influence the electronic properties of the bilayer system. 

Furthermore, we apply a commensurate twist angle of $36.87^\circ$ to the three stackings and observe that each undergoes a topological phase transition after twisting. Specifically, AB and AA stackings transform into $d$-wave altermagnetic bipolarized Weyl semimetals, while AC stacking becomes an unconventional compensated magnetic Weyl semimetal. A key common feature is the existence of eight spin-polarized Weyl points protected by $\left\{ C_2 T||C_{2z} T\right\}$ spin symmetry at the four corners of the BZ, with contributions from Cr orbitals originating from different layers. These Weyl points are well separated within the BZ and located near the Fermi level, making them excellent candidates for future experimental studies of low-energy Weyl fermions. The twist angle causes each Dirac point on the original high-symmetry line to split into two Weyl points and adiabatically shift their positions inside the BZ. Moreover, this symmetry is still preserved when other commensurate twist angles are applied. That is to say, the Weyl points will still exist. This demonstrates that the twist angle can act as a control knob to design and manipulate MTPTs. Experimentally, tear-and-stack assembly and atomic force microscope
manipulation enable twist-angle control with an accuracy of approximately
$0.1^\circ$ \cite{du2021engineering}. Cutting-rotation-stacking
\cite{cutting}, mechanical bending \cite{mechanical}, electrostatic
microelectromechanical systems (MEMS) \cite{MEMS}, and the recently proposed
all-optical approach \cite{OPTICAL} provide complementary routes for precise
and in situ control of the twist angle. These advances suggest that the
twist-driven magnetic topological phase transition proposed here may be
experimentally accessible in principle. Our findings highlight the profound impact of twist angles on the topological properties of stacked bilayer CrO. This work not only provides a versatile platform for exploring altermagnetism and topological phenomena in quantum materials, but also paves the way for potential applications in spintronics and twistronics.

\begin{acknowledgments}
This work was financially supported by the National Natural Science Foundation of China (Grant No.~12434009, No.~62476278 and No.~12174443), the National Key R$\&$D Program of China (Grant No.~2024YFA1408601), the Fundamental Research Funds for the Central Universities, and the Research Funds of Renmin University of China (Grant No.~24XNKJ15). Computational resources have been provided by the Physical Laboratory of High Performance Computing at Renmin University of China.
\end{acknowledgments}

\nocite{*}

\bibliography{Reference}

\end{document}